\documentclass[twocolumn,english,showpacs]{revtex4}
\usepackage{babel,amsmath,amssymb,dcolumn}
\usepackage[dvips]{graphics}

\begin{document}

\title{The ringing wormholes}

\author{R. A. Konoplya}
\email{konoplya@fma.if.usp.br}
\affiliation{Instituto de F\'{\i}sica, Universidade de S\~{a}o Paulo \\
C.P. 66318, 05315-970, S\~{a}o Paulo-SP, Brazil}

\author{C. Molina}
\email{cmolina@usp.br}
\affiliation{Escola de Artes, Ci\^{e}ncias e Humanidades, Universidade de
  S\~{a}o Paulo\\ Av. Arlindo Bettio 1000, CEP 05315, S\~{a}o
  Paulo-SP, Brazil}

%
%
%

\begin{abstract}
We investigate the response of the traversable wormholes to the
external perturbations through finding their characteristic
frequencies and time-domain profiles. The considered solution
describes traversable wormholes between the branes in the two brane
Randall-Sundrum model and was previously found within Einstein gravity
with a conformally coupled scalar field. The evolution of
perturbations of a wormhole is similar to that of a black hole and
represents damped oscillations (ringing) at intermediately late times,
which are suppressed by power law tails (proportional to $t^{-2}$ for
monopole perturbations) at asymptotically late times.    
\end{abstract}

\maketitle

\section{Introduction}

Since the seminal paper of Moriss and Torne \cite{1} where it was
shown that the wormholes with the two mouths and a throat can exist in
the nature but require the specific ``exotic'' matter to support them,
wormholes attracted considerable interest in the realm of classical
and quantum gravity \cite{2}. The most interesting kind of  wormholes,
called traversable wormholes, allows traveling through their throats,
and thereby makes it possible to span large distances or even travel
to other Universes \cite{1}. Yet, wormholes cannot be formed as a
result of a star's collapse, and, to construct the traversable
wormhole one has to violate null energy condition, the weakest of the
conditions a ``good'' classical system should obey. This condition,
however, are known to be violated for quantum systems and even for
scalar field non-minimally coupled to gravity \cite{3}. A unified view
on wormholes says that wormholes are objects in the same line as stars
and black holes: stars are made from normal matter, black holes are
``made'' of vacuum, wormholes are made of exotic matter. That is why
sometimes the wormholes are called ``exotic stars''
\cite{4}. Considerable interest to traversable wormholes was
stipulated also by possibility of closed time-like curves, i.e. time
machines in the wormhole background. Thus different formulations of
chronological protection were made within the above context.

If one considers the possibility of existence of wormholes  seriously,
one should investigate such properties of wormholes as stability, 
interaction with external fields, and, response of wormholes to
external perturbations. If they exist, the wormholes should produce
micro-lensing effect on point sources \cite{5}, and very large
wormholes can produce even macro-lensing effect \cite{6}. 

Recently, the traversable wormholes have gained interest due to
possible interpretation within the extra dimensional Randall-Sundrum
scenario \cite{7}. That is the  simplest brane-world scenario with a warped
bulk geometry, the two brane model, where a pair of
four dimensional branes live in a five-dimensional bulk with 
negative cosmological constant. 
In the four-dimensional effective theory approach \cite{8}, 
the traversable wormholes connect the branes and the wormhole solution
can be obtained as a 4-dimensional effective solution with
non-trivial radion \cite{9}. The above wormhole solution was first obtained
in \cite{10} within the Einstein gravity with a conformally coupled
scalar field and analyzed in \cite{3}.
Its stability against monopole perturbations has been recently proved
in \cite{9}. Yet, as far as we are aware,  the response of a
wormhole to external perturbations was not investigated so far.

The response of a stable wormhole should represent some damping oscillating 
signal which can be decomposed, with Laplace transformation techniques,
into a set of modes, at least at late times. We shall show in the
present paper that these modes are similar to the (quasi)normal
modes of black holes \cite{11}, which have been extensively
investigated recently in classical gravity and string theory
\cite{12}. The quasinormal modes (QNM) of black holes are important
because they dominate in the intermediately late time decay of a
perturbation  and do not depend upon the way they were
excited. Quasinormal modes of black holes depend only on the
parameters of a black hole and are called therefore the ``footprints''
of a black hole (for more recent references on black hole QNMs see
\cite{13}). They are called quasinormal because they are analog of
normal modes of a closed system. The black hole, and, in the same
fashion, the wormhole is an open system: the damping gravitational
waves induced by initial perturbations give rise to the lost of energy
of a black or wormholes. The time-independent problem for perturbations of
a wormhole turned out to be quite similar to that for a black hole:
one has to find the solutions of the wave-like equations which satisfy
the appropriate boundary conditions, which we shall discuss further in
details.  

In this paper we considered the decay of monopole perturbations
of the traversable wormhole solution which naturally appears 
in the effective theory approach of the Randall-Sundrum two brane
model and in the Einstein theory with non-minimally coupled scalar
field. We find the quasinormal frequencies which govern the decay
of the perturbations at intermediately late times by three different
approaches: P\"{o}schl-Teller approximation, WKB approach, and by
characteristic integration. All three approaches show good
agreement. In addition we found the profile of a response signal in
the time domain and observed that at asymptotically late times, when
quasinormal modes are suppressed by the so-called ``tails'', the
perturbations decay according to the law  $\propto t^{-2}$ as $t$
approaches infinity.

\section{Wormhole perturbations}   

The traversable wormhole we shall consider here is described by the
one-parameter family of solutions:
\begin{equation}
d s^2 = p \left[ - d t^2 + \frac{M^2 (1+y^4)}{y^4} \left(d y^2 + y^2 d
\Omega_{2}^{2}\right) \right]
\label{metric}
\end{equation}
where the scalar field $\Phi$ obeys
\begin{equation}
\Phi = \frac{1}{p} \left(\frac{1-y}{1+y}\right).
\label{scalar_field}
\end{equation}
This solution appears when dealing with the Einstein-Hilbert action
of the form 
\begin{equation} 
S = m_{p}^{2} \int d^{4} x \sqrt{-g} \Phi R,
\end{equation}
thereby implying the equations of motion:
\begin{equation}  
\Phi R_{\mu \nu} = \nabla_{\mu} \nabla_{\nu} \Phi, 
\end{equation}
\begin{equation} 
\Box \Phi = 0.
\end{equation}
The metric (\ref{metric}) can be interpreted as the induced metric on the
brane when considering the tensionless vacuum branes compactified
on an $S^{1}/Z_{2}$ orbiforld, which implies the following bulk
Ansatz: 
\begin{equation}
d s^2 = g_{\mu \nu} (x) d x^{\mu} d x^{\nu} + \Phi^{2} (x) d Y^{2}.  
\end{equation}
In this case the bulk warp is absent. When considering the branes with
tension, due to the presence of the bulk warp, the induced metric on
the positive tension and negative tension branes $g_{\mu \nu}^{1}$, $
g_{\mu \nu}^{2}$  are no longer identical:
\begin{equation}
g_{\mu \nu}^{1} = \frac{1}{4} (1 + \Phi)^{2} g_{\mu \nu} \, , \, 
g_{\mu \nu}^{2} = \frac{1}{4} (1 - \Phi)^{2} g_{\mu \nu} 
\end{equation}

\begin{figure}
\resizebox{0.9\linewidth}{!}{\includegraphics*{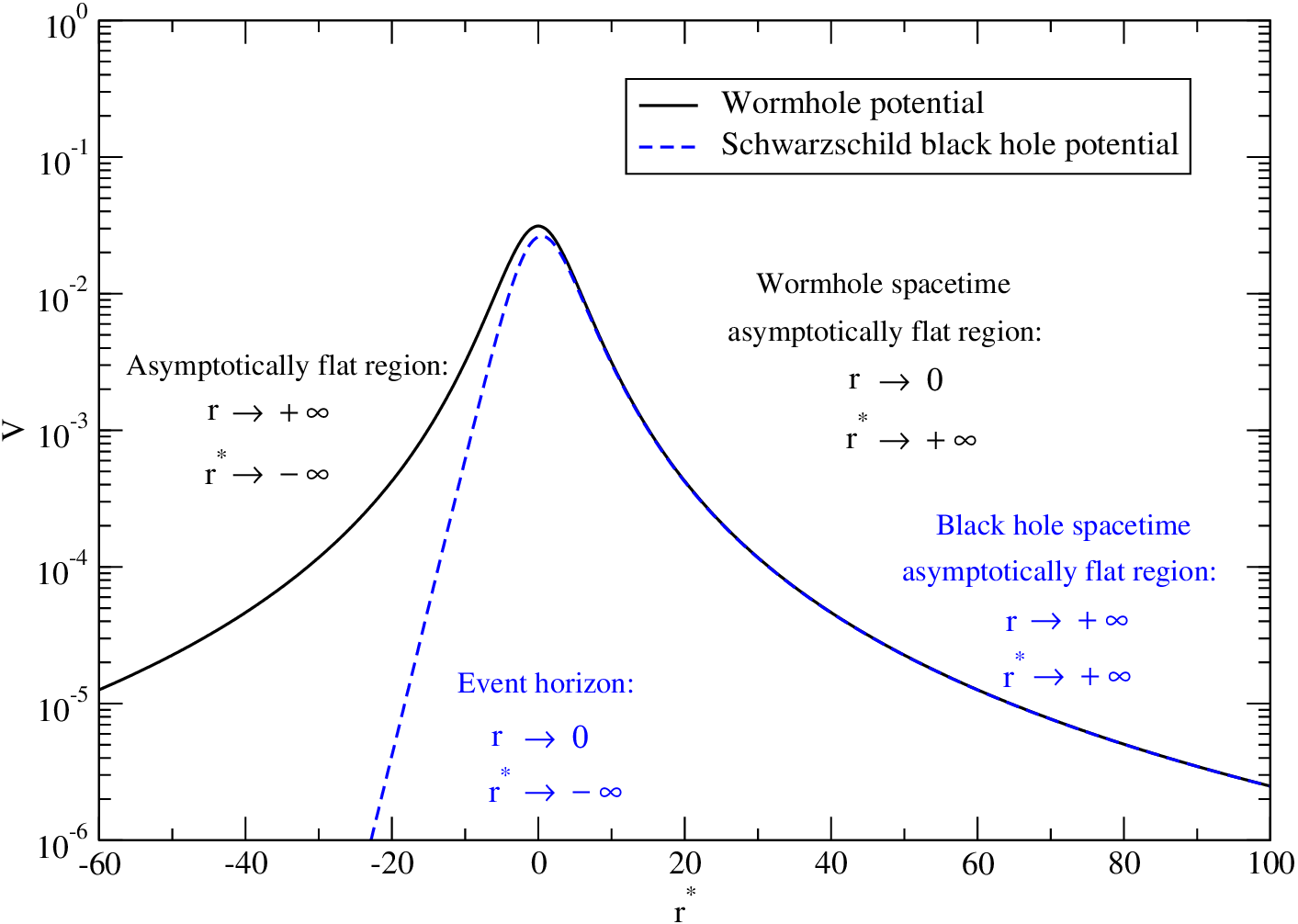}}
\caption{Semi-log graph of the wormhole and Schwarzschild black hole
  effective potentials as a function of the tortoise coordinate $r^{*}$.}     
\label{fig_potential}
\end{figure}

The linear perturbations of the background metric (\ref{metric}) and
of the scalar field (\ref{scalar_field}), generally have the form:
\begin{equation}
g_{\mu \nu}^{'} (y, t) = g_{\mu \nu} (y) + \delta  g_{\mu \nu} (y, t)
\end{equation}
\begin{equation}
\Phi^{'} (y, t) = \Phi (y) + \delta  \Phi (y, t)
\end{equation}
In \cite{9} it was shown that it is appropriate to analyze the perturbations
in the gauge where the scalar field perturbations decouple from that
of the metric, namely to take
\begin{equation}
\delta (\sqrt{-g} g^{yy}) =0, \quad \delta g_{yt} = 0. 
\end{equation}
Note that according to spherical symmetry $ \delta g_{y \theta} =
\delta g_{y \phi} =0 $. After simple algebra  one can reduce the
perturbation equation to the wave-like form:
\begin{equation}
\left[ \frac{d^2}{dr^{*2}} + \omega^2 - V(r^*)\right] \Psi = 0. 
\label{time_indep_eq}
\end{equation}
where the tortoise coordinate is
\begin{equation}
r^{*} = M \left[ \frac{1}{y} -y - 2 \ln y \right],
\end{equation}
and the effective potential has the form \cite{9}:
\begin{equation} 
V = \frac{2 y^3}{ M^{2} (1+y)^{6}}.
\end{equation}
This effective potential, similarly to the black hole potential (see
Fig. \ref{fig_potential}) has the form of the potential barrier
which approaches constant asymptotic values at both infinities. Note
also that the monopole perturbations of the above wormhole were previously
considered in \cite{Grinyok} in the gauge where $g_{\theta \theta} = 0$,
and the scalar field perturbations cannot be decoupled from those of the
metric. In that gauge, contrary to the suggestion of \cite{Grinyok} 
about possibility of infinitely growing modes, 
the consideration in \cite{9} shows that the metric 
is stable and that the gauge $g_{\theta \theta} = 0$ is not suitable 
for stability analysis. Note also, that the effective potential does
not depend on the parameter $p$ and, thereby, the stability analysis in
\cite{9} is valid, in particular, for the case $p=1$, corresponding to
the so-called BBMB black hole \cite{20}.

\section{Quasinormal modes and late time tails}

Let us start from definition of quasinormal modes for black holes in
asymptotically flat space-time. Quasinormal modes of black holes are
defined as the solutions of the time-dependent version of
Eq. (\ref{time_indep_eq}) which satisfy both boundary conditions which
requires the purely out-going waves at spatial infinity and purely
in-going waves at the event horizon. These modes are indexed by a set
of (usually) complex frequencies  $\omega$. The requirement of purely
out-going waves at spatial infinity is natural: at asymptotically far
and flat region a passive observer does not emit waves
\cite{Starina}. Another interpretation of quasinormal (QN) boundary
conditions at infinity is the requirement of resonance for $\omega$,
i.e. the requirement of divergence of the reflection amplitude
\cite{Will}. Also it is natural to forbid waves coming from the event
horizon.  The QN boundary conditions may also change if a black hole
is immersed in some cosmological background, such as de Sitter
\cite{14}, or anti-de Sitter \cite{15}. Thus, for example, for an
asymptotically anti-de Sitter black holes, the effective potential is
divergent at infinity, and the Dirichlet boundary conditions are more
appropriate \cite{15a}.

\begin{figure}
\resizebox{0.9\linewidth}{!}{\includegraphics*{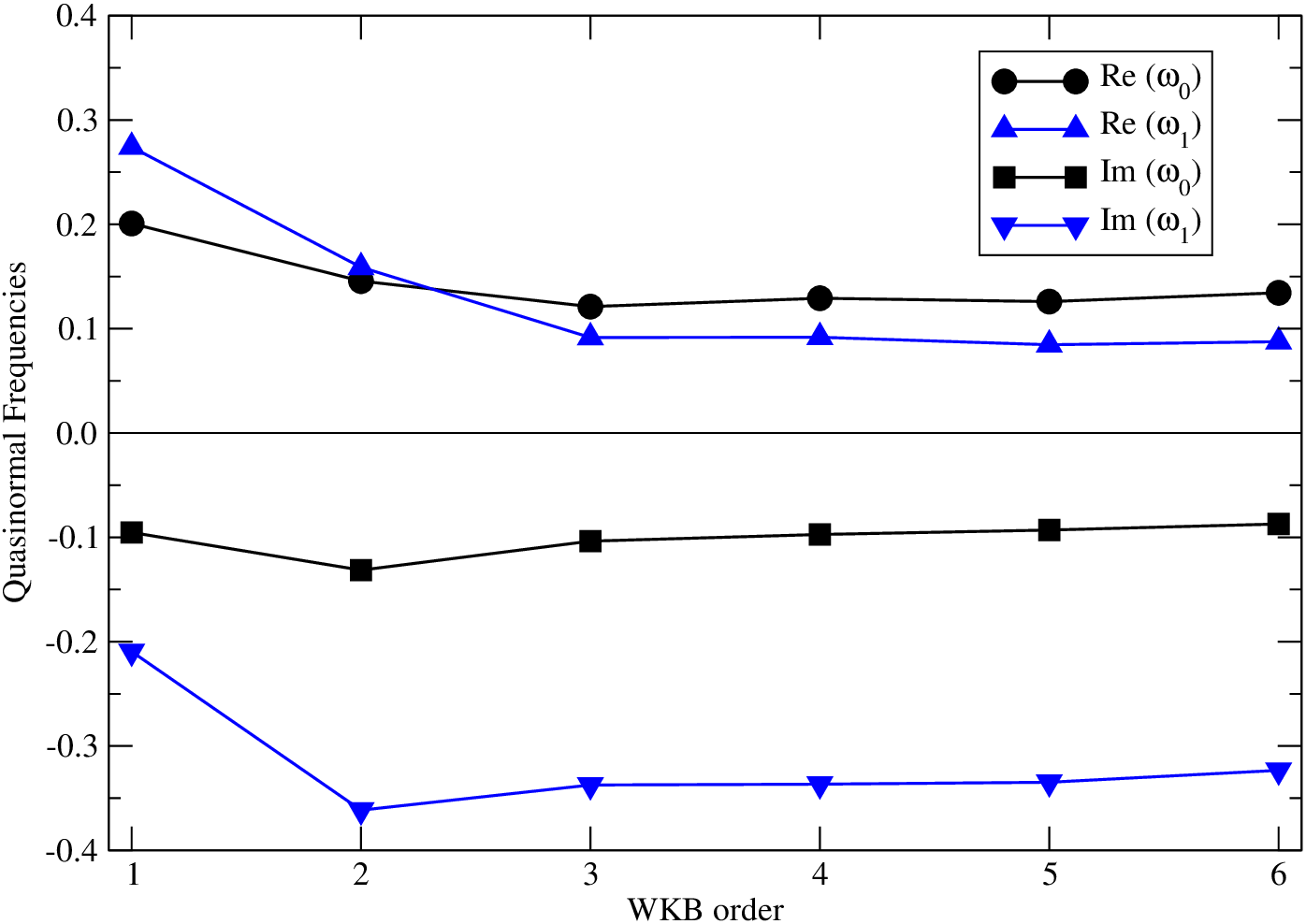}}
\caption{Graph showing reasonable convergence of the WKB method with
  its order, for the fundamental mode ($\omega_{0}$) and first
  overtone ($\omega_{1}$).}      
\label{fig_WKB}
\end{figure}

For the wormhole we have qualitatively different situation, yet 
leading to the same boundary conditions as those for black holes. 
The effective potential with respect to the tortoise coordinate, being
a potential barrier which approaches constant values at both
infinities $r^{*}=-\infty$  and $r^{*}=+\infty$, but has no longer
exponential asymptotic at $r^{*}=-\infty$ as it takes place for black
holes (see Fig. \ref{fig_potential}). Note however that now ``minus
infinity'' corresponds to ``our'' spatial infinity and ``plus
infinity'' corresponds to the other asymptotically flat region which
is separated from our  space-time by a bridge. The natural QN boundary
conditions for both infinities are the requirement of purely out-going
waves into both infinities. Thus no waves are coming from both
asymptotically flat regions:  
\begin{equation} 
\Psi \approx e^{i \omega \pm r^{*}}, \quad r^{*} \rightarrow \pm \infty.
\end{equation}
Note that $\omega = \textrm{Re} \left(\omega \right) + i \,
\textrm{Im} \left( \omega \right)$ with $\textrm{Im}<0$, and
$\textrm{Re} \left( \omega \right)$ is chosen to be positive.

The above boundary conditions for wormhole make it possible to use 
the WKB \cite{16} and P\"{o}schl-Teller approaches \cite{17}
previously developed and used for black holes \cite{18}. Thus we find
the quasinormal modes of the wormhole within three different
approximations: P\"{o}schl-Teller formula, WKB formula in the 6th
order beyond the eikonal approximation \cite{19} and in all previous
orders (see Fig. \ref{fig_WKB}), and characteristic integration
\cite{integration}. All approaches shows good agreement for the fundamental
overtone which is the dominating in the signal. That is why the
frequency found through the characteristic integration of the response
drown in time-domain is so close to the value for the fundamental
overtone obtained in the frequency domain (see Table \ref{QNM}).

\begin{table}
\caption{Values for the quasinormal frequencies for the fundamental
  mode $\omega_{0}$ and first overtone $\omega{1}$, obtained from
  sixth order WKB method, P\"{o}schl-Teller approximation and directly
  from time-domain profile.}  
\label{QNM}
\begin{ruledtabular}
\begin{tabular}{lcccc}
\multicolumn{1}{l}{  }        &
\multicolumn{2}{c}{Fundamental Mode}&
\multicolumn{2}{c}{First Overtone}\\
  & Re($\omega_0$) & -Im($\omega_0$) &
    Re($\omega_1$) & -Im($\omega_1$)    \\
\hline \\ 
WKB (6th order)   & 0.1344 & 0.08717 & 0.08759 & 0.323312  \\
P\"{o}schl-Teller & 0.1398 & 0.10825 & 0.13975 & 0.32476   \\ 
Time evolution    & 0.1323 & 0.09760 & -       & -         \\ 
\end{tabular} 
\end{ruledtabular}
\end{table}


If one would like to compare the QN modes of a  wormhole with those 
of a black hole, then the immediate question is: which model of a black
hole is adequate for such a comparison? The black hole coming from
the same solution (\ref{metric}) when $p=1$ probably is not a good
candidate, since corresponding black hole \cite{20} has the topology of
the extremal Reissner-Nordstr\"{o}m black hole, thereby representing
very specific case. Moreover, monopole perturbations of such black
hole are governed by the effective potential Eq. (49) of \cite{9}, which 
is isospectral to the effective potential for the considered 
here wormhole (we checked also numerically that both potentials
correspond to the same QN modes). Nevertheless, one should distinguish
the quasinormal spectrum of the extremal Reissner-Nordstr\"{o}m black
hole \cite{20a} from that discussed here, since in the present case 
we have the new type of perturbations: the perturbed scalar field
non-minimally coupled to gravity. This, certainly totally changes
the QN spectrum.      

Alternative simple choice for comparison of a black hole and a
wormhole within spherical symmetry, is, ordinary Schwarzschild black
hole. From the above data (see Fig. \ref{fig_potential},
Fig. \ref{fig_WKB} and Table \ref{QNM}), one can learn that the
quasinormal modes of a wormhole is considerably different from those
of the black hole \cite{Will}.  The wormhole QN modes have much
greater oscillation frequency under the damping rate of the same
order, thus the ratio $\textrm{Re} \left(\omega\right) / \textrm{Im}
\left( \omega \right)$ is larger for a wormhole, that is, the
considered  wormhole is much better oscillator then a  Schwartzschild
black hole of the same mass. It is certainly unclear whether this
feature of the wormhole is stipulated by the specific type of matter
supporting the wormhole (scalar field), or is generic for wormholes. 

\begin{figure}
\resizebox{0.9\linewidth}{!}{\includegraphics*{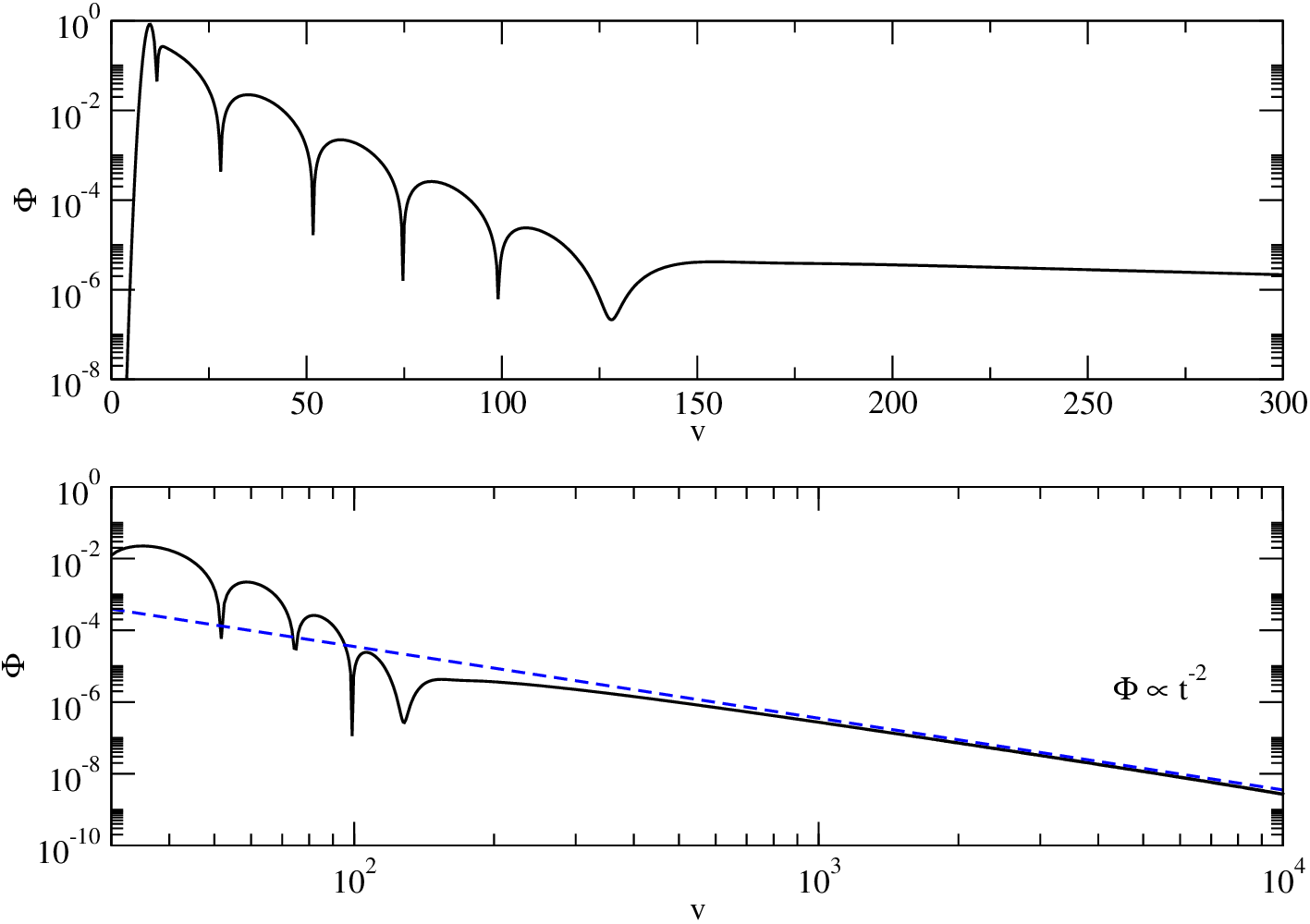}}
\caption{(above) Time-domain profile (semi-log graph) of the scalar
  field perturbation, showing quasinormal ringing and late time
  tail. The fundamental quasinormal frequency calculated from the
  numerical data is \mbox{$\omega_{0} = 0.1323 - i 9.760 \times
  10^{-2}$}. (below) Log-log graph of the late time power law tail
  proportional to $t^{-2}$, as indicated by the dashed line.  }       
\label{fig_decay}
\end{figure}

At asymptotically late times the perturbation decays
according to a power law: as can be seen from Fig. \ref{fig_decay} the
late time tail is obeying the low: $\propto t^{-2}$. This is different
from the Schwartzschild black hole late time behavior, where the power
law for the $\ell$-th multipole is proportional to $t^{-2 \ell +3}$,
i.e. $\propto t^{-1}$ for the dominating gravitational multipole, and
$\propto t^{-3}$ for the monopole scalar field perturbations.

Note that for metric  perturbations of the Schwartzschild black hole,
the first (and the second)  multipole is not dynamical and corresponds
to the infinitesimal change of mass. It is well understood then, that
the striking difference in perturbations a black hole and a wormhole is
that the monopole perturbations of a wormhole become dynamical, and
since they correspond to the lowest multipole index, one can suggest
that they will dominate over contributions of higher multipoles in a
wormhole quasinormal ringing, as it takes place for black holes.

\section{Conclusion}

In the present paper the evolution of perturbations of a
wormhole space-time has been investigated for the first time.
We have shown that the evolution of a wormhole response 
has the same stages as that of a black hole, i.e. includes
the quasinormal ringing and power-law asymptotically late time tails.
We suggest that these main features of the evolution of the wormhole
response, are not stipulated by specific matter supporting wormholes
but are generic for different wormholes.   
We were limited here by a monopole perturbations 
and consideration of higher multipole perturbations seems to be 
promising problem. Also, consideration of the influence of quantum
effects, like Hawking radiation, onto classical (quasinormal)
radiation \cite{21} is beyond the scope of this paper.


\begin{acknowledgments}
 This work was partially supported by \emph{Funda\c{c}\~{a}o de Amparo
\`{a} Pesquisa do Estado de S\~{a}o Paulo (FAPESP)}, Brazil.
\end{acknowledgments}


%
%
\end{document}